# A Genealogy for Finite Kneading Sequences of Bimodal Maps on the Interval

John Ringland [1] and Charles Tresser [2]

## Abstract

We generate all the finite kneading sequences of one of the two kinds of bimodal map on the interval, building each sequence uniquely from a pair of shorter ones. There is a single pair at generation 0, with members of length 1. Concomitant with this genealogy of kneading sequences is a unified genealogy of all the periodic orbits.

(1) Math. Dept., SUNY Buffalo, Buffalo, NY 14214.
(2) I.B.M., PO Box 218, Yorktown Heights, NY 10598.

# I. Introduction

A continuous map f:[0,1]→[0,1] is {+-+}-bimodal if there exists two points c, k with $0 < c < k < 1$ such that f is increasing on [0,c]∪[k,1], and decreasing on [c,k].  If f is constant in some interval containing c, c is chosen as the middle of this interval, and the same convention holds for k. The points c and k are called the turning points of the map.

Following [MT], we code the dynamics of such a map in the following way.  Given a {+-+}-bimodal map, we define:

- the address  a(x) of x∈ [0,1] as

$$\begin{array}{l} \text{L if } x \in [0,c), \\ \text{C if } x = c, \\ \text{M if } x \in (c,k), \\ \text{K if } x = k, \\ \text{R if } x \in (k,1], \end{array}$$

- the extended itinerary of x as the word

$$I_e(x) = a(x)\, a(f(x))\, a(f^2(x))\, ...$$

in the alphabet $\mathcal{A}$ ={L,C,M,K,R}, where

$$f^0 = \text{Id}, \quad f^n = f \circ f^{n-1} \text{ for } n>0,$$

- the itinerary of x as the word $I_e(x)$ if it does not contain any C or K, and as the finite word obtained from $I_e(x)$ by cutting after the first symbol C or K if any,

- and the kneading data  of the map f as the pair

$$K(f)=(K^+(f),K^-(f))=(I(f(c)),I(f(k))),$$

where the special itineraries $K^+(f)$ and $K^-(f)$ are called the kneading sequences of f.

What we do here is generate hierarchically all the finite kneading sequences realizable by {+-+}-bimodal maps in a natural way that  (i) is related to local bifurcation (Theorems 1 and 2), (ii) allows for a direct comparison of entropies (Theorem 3), and (iii) provides a way to link different dynamical behavior (Theorem 4). Our result can be compared to the generation of all itineraries for rotations using the Farey tree (see e.g. [STZ] and references therein to an abundant previous literature on this related subject) or the decomposition of kneading sequences for Lorenz-like maps [TW]. It is also to be compared with a previous genealogical result for unimodal maps [D].

The paper is organized as follows. In section II, we present a symbolic space $\mathcal{V}$, directly motivated by the kneading theory of {+-+}-bimodal maps, whose finite words are all possible finite kneading sequences of such maps, and we formulate our main results about the space $\mathcal{V}$. Proofs are provided in section III. The main result in section II can be interpreted as a hierarchical organization of all the finite kneading sequences of {+-+}-bimodal maps, giving in particular a new partial order on these sequences: this will be discussed in section IV in connection with what we call the topological entropy of a kneading sequence. In section V, we use stunted {+-+}-saw-tooth maps to display these results graphically: this two-parameter family is simple enough to allow for a fairly complete description of the subset **S** of the parameter space corresponding to maps with (at least) one finite kneading sequence. Hence one can draw pictures which are provably topologically exact. Taking some quotient of the structure so generated for the stunted {+-+}-saw-tooth maps yields the conjectural picture for the counterpart to **S** in families of smooth maps, proposed in [RS1] and [RS2] on the basis of local bifurcation diagrams. A small subset of the conjectural picture, computed with an explicit family of cubic maps, is presented in Figure 0 for orientation. (More fully described figures appear in the last section.) Early pictures of this kind appeared in [FK] and [PG]: we refer the reader to [MaT1-MaT3], [RS1], [RS2] and the more recent paper [Mi] for more complete references.

The {+-+}-bimodal maps are often studied because they serve to describe the dynamics of the simplest non-monotonic degree-one circle maps (see e.g. [B] and [MaT1]). There is another kind of bimodal map, the {-+-}-bimodal maps, which are decreasing on the first interval of monotonocity. The {-+-}-bimodal maps have a somewhat more complicated bifurcation diagram, as suggested by some partial results in [MaT2, MaT3], and a description of this class of maps in parallel to what is done here will be offered in a forthcoming paper [R].

## II. Statement of the symbolic results

Sections II and III constitute a self-contained symbolic discussion, and do not rely on any properties of interval maps. However, the theorems developed in these sections will later be interpreted as providing a genealogy for the kneading sequences of {+-+}-bimodal maps, so we do mention, as we go, some motivations in the theory of such maps.

### II.1 Preliminary Definitions

We first make enough definitions to allow us to state the main results. Following standard usage, we define $\mathcal{A}^*$ to be the set of empty or finite words written with alphabet $\mathcal{A}$ = {L,C,M,K,R}, and $\underline{\mathcal{A}}$ to be the set of infinite words written with alphabet $\mathcal{A}$. We further define $\mathcal{W}$ to be the set of all words in $\underline{\mathcal{A}}$ truncated after the first C or K if any, and $\mathcal{W}^\#$ to be the set of finite words in $\mathcal{W}$. We remark that $\mathcal{W}$ contains all kneading sequences of bimodal maps, and $\mathcal{W}^\#$ all the finite kneading sequences. In order to discuss subwords of words in $\mathcal{W}^\#$ we introduce the reduced alphabet $\mathcal{B}$ = {L,M,R} and define $\mathcal{B}^*$ as the set of empty or finite words written with alphabet $\mathcal{B}$. Note that each word in $\mathcal{W}^\#$ is of the form BQ with B $\in \mathcal{B}^*$ and Q $\in$ {C,K}.

In what follows, we will use upper case letters not in $\mathcal{A}$ to denote words in the various sets defined above. For F $\in \mathcal{A}^*$, G $\in \mathcal{A}^* \cup \underline{\mathcal{A}}$, we write FG to mean the concatenation of the words F and G. (FG $\in \mathcal{A}^* \cup \underline{\mathcal{A}}$). We use 1 to denote the empty word: for G $\in \mathcal{A}^* \cup \underline{\mathcal{A}}$, 1G=G1=G. For F $\in \mathcal{A}^*$, we use $[F]^n$ to mean the concatenation of n copies of F: $[F]^0 = 1$, and for n>0, $[F]^n = F[F]^{n-1}$. $|W|$ stands for the length of (number of letters in) W $\in \mathcal{A}^*$.

We define an ordering $\leq$ on $\mathcal{W}$ by first imposing the following ordering on $\mathcal{A}$:

$$L < C < M < K < R$$

To induce an ordering on $\mathcal{W}$, we define the parity of a word W in $\mathcal{B}^*$ as even if it contains an even number of M's, and odd otherwise, and a parity function $\rho: \mathcal{B}^* \to \{-1,1\}$ by

$$\rho(W) = \begin{cases} +1 \text{ if W is even} \\ -1 \text{ if W is odd.} \end{cases}$$

Then, if two distinct words V and W in $\mathcal{W}$ are written as V=TAX and W=TBY, with T∈ $\mathcal{B}^*$, A and B∈ $\mathcal{A}$ and A<B, and X and Y ∈ $\mathcal{W}$,

    V < W        if        ρ(T) = +1,

    W < V        if        ρ(T) = –1.

We remark that for any {+–+}-bimodal map on [0,1] and any x,y in [0,1], we have $x < y \Rightarrow I(x) \leq I(y)$ and $I(x) < I(y) \Rightarrow x < y$ (see e.g. [MT]).

In order to specify the words in $\mathcal{W}$ that are realizable as the kneading sequence of a bimodal map, we introduce the <u>shift map</u> $\sigma: \mathcal{A}^* \cup \underline{\mathcal{A}} \to \mathcal{A}^* \cup \underline{\mathcal{A}}$, defined by: $\sigma(W) = 1$ if $|W| \leq 1$, and, for $|W| > 1$, writing W=AW′, with |A|=1, $\sigma(W) = W'$. We will use the notation $\sigma^0(W) = W$, and $\sigma^k(W) = \sigma(\sigma^{k-1}(W))$, k>0. A word W in $\mathcal{W}$ will be called <u>minimal</u> if it bounds all its shifts from below ($W \leq \sigma^k(W) \ \forall \ k \in \{0,1,...|W|\}$), <u>maximal</u> if it bounds all its shifts from above ($\sigma^k(W) \leq W \ \forall \ k \in \{0,1,...|W|\}$), and <u>extremal</u> if it is minimal or maximal. We remark that being extremal is a necessary and sufficient condition that the word occur as a kneading sequence of some {+-+}-bimodal map: necessity is rather obvious, and a proof of sufficiency is easily provided using stunted sawtooth families. We are thus motivated to define $\mathcal{V}$ as the set of all extremal words in $\mathcal{W}$, and $\mathcal{V}^\#$ as the set of finite words in $\mathcal{V}$.

A pair, (V,W), of words in $\mathcal{V}$ will be called <u>admissible</u> iff all shifts of V and W are bounded from above by V and from below by W: that is, iff $\forall \ i \in \{0,1,...|W|\}, W \leq \sigma^i(W) < V$ and $\forall \ j \in \{0,1,...|V|\}, W < \sigma^j(V) \leq V$. As motivation we point out that admissibility is a necessary and sufficient condition that the pair appear as the kneading data of some {+-+}-bimodal map.

In the following, we will frequently be concerned with the replacement in a word of C by L or M, and of K by M or R. We define $C^{-1} = L$, $C^{+1} = M$, $K^{-1} = M$, $K^{+1} = R$. Note that $C^{-1} < C < C^{+1}$ and $K^{-1} < K < K^{+1}$. This exponential notation will frequently be used in conjunction with the parity function. Observe that $AC^{-\rho(A)} < AC < AC^{+\rho(A)}$ and $AK^{-\rho(A)} < AK < AK^{+\rho(A)}$.

## II.2  Hierarchical generation of all words in $\mathcal{V}^{\#}$

**Definitions:** Among finite admissible pairs we single out three classes.

An admissible pair of words in $\mathcal{V}^{\#} \times \mathcal{V}^{\#}$ of the form (AQ,BQ), Q$\in$ {C,K}, will be called a $\underline{\psi\text{-seed}}$.

An admissible pair of words in $\mathcal{V}^{\#} \times \mathcal{V}^{\#}$ of the form (AK,BC) will be called a $\underline{\chi\text{-seed}}$.

In the sequel, $\mathcal{V}^{\#} \times \mathcal{V}^{\#}$ will be understood as the quotient of the bare product by the relation (C,C)$\approx$(K,K); (C,C) and (K,K) will be considered as two equivalent representations of the $\underline{\alpha\text{-seed}}$.

**Remarks:** The $\psi$-seed corresponds to the kneading data of a map with one critical point periodic and in the orbit of the other. The $\chi$-seed corresponds to the kneading data of a map with both critical points belonging to the same periodic orbit. The $\alpha$-seed is thought of as the kneading data of a map with a 1-periodic cubic-like inflection point. The class of finite admissible pairs *not* mentioned above are those of the form (AC,BK), which correspond to the kneading data of a map whose critical points belong to two different periodic orbits. They are not relevant to our present purposes for they do not participate in the hierarchy: AC and BK are the only finite kneading sequences that exist for smooth maps arbitrarily close to one with kneading data (AC,BK).

**Theorem 1:**

$\alpha$. The words
$$\alpha_{+n} \equiv [R]^{n-1}K, \quad n \in \{1,2,...\},$$
are maximal, and the words
$$\alpha_{-n} \equiv [L]^{n-1}C, \quad n \in \{1,2,...\},$$
are minimal.
We say that these words emanate from the $\alpha$-seed.

$\psi$. If (AK,BK) is a $\psi$-seed, then the words
$$\Psi_{\pm n}(AK,BK) \equiv AK^{\pm\rho(A)}[BK^{\rho(B)}]^{n-1}BK, \quad n \in \{1,2,...\},$$
are maximal.
If (AC,BC) is a $\psi$-seed, then the words
$$\Psi_{\pm n}(AC,BC) \equiv BC^{\pm\rho(B)}[AC^{-\rho(A)}]^{n-1}AC, \quad n \in \{1,2,...\},$$
are minimal.
We say that the words $\Psi_{\pm n}(AQ,BQ)$ emanate from the $\psi$-seed (AQ,BQ).

χ. If (AK,BC) is a χ-seed, then the words
$$\beta_{(+1,\pm 1)}(AK,BC) \equiv AK^{\pm\rho(A)}BC$$
$$\chi_{(+1,\pm n)}(AK,BC) \equiv AK^{+\rho(A)}[BK^{\pm\rho(B)}AK^{-(\pm\rho(A))}]^{n-1}BC^{-\rho(B)}AK, \quad n \in \{1,2,...\},$$
are maximal, and the words
$$\beta_{(-1,\pm 1)}(AK,BC) \equiv BC^{\pm\rho(B)}AK$$
$$\chi_{(-1,\pm n)}(AK,BC) \equiv BC^{-\rho(B)}[AK^{-(\pm\rho(A))}BC^{\pm\rho(B)}]^{n-1}AK^{-(\pm\rho(A))}BC, \quad n \in \{1,2,...\},$$
are minimal.

We say that these words emanate from the χ-seed (AK,BC).

**Remark:** The word emanation is used because in smooth bimodal families an emanating word exists as a kneading sequence of the map on a curve in the two-dimensional parameter space that abuts on the point where the seed from which the sequence emanates is the kneading data [RS2].

We now state the main result of the paper.

**Theorem 2:** Every word in $\mathcal{V}^{\#}$ emanates from a unique seed.

## III. Proof of Theorems 1 and 2

Theorems 1 and 2 are proved using a substantial number of preparatory lemmas. The proofs of these lemmas are rather mechanical, and involve much near-repetition. For this reason, we state most of them without proof. Exceptions are Lemma B which is fundamental, being used in the proof of nearly every other statement, and Lemma C3a whose proof is included to illustrate the character of the other proofs. The proofs of Theorems 1 and 2, using the lemmas, are given in full.

In addition to the terminology introduced in section II, we will use the following. We define the <u>truncations</u>, $\tau_k(W)$, of a word W in $\mathcal{A}^* \cup \underline{\mathcal{A}}$, k in $\{0,1,...\}$, by $W = \tau_k(W)\sigma^k(W)$. Note that $\tau_0(W) = 1$, and $\tau_k(W) = W$ for $k > |W|$. Thus $\sigma^k(W)$ is the word obtained from W by deleting its first k letters, and $\tau_k(W)$ is the word consisting of the first k letters of W. Note that for all k in $\{0,...,|W|\}$, $|\sigma^k(W)| = |W|-k$, and $|\tau_k(W)| = k$. We define an <u>order function</u>, $O: \mathcal{W} \times \mathcal{W} \to \{-1,0,1\}$, by

$$O(V,W) = \begin{cases} +1 & \text{if } V < W, \\ 0 & \text{if } V=W, \\ -1 & \text{if } W < V. \end{cases}$$

Note that for all (V,W) in $\mathcal{V}^{\#} \times \mathcal{V}^{\#}$, $O(W,V) = -O(V,W)$, and if $\tau_k(V) = \tau_k(W)$ then $O(V,W) = \rho(\tau_k(V)) \, O(\sigma^k(V), \sigma^k(W))$. We define for W in $\mathcal{A}^*$, $\mathcal{N}(W)$ as the position of W in the list ordered by < of all words in $\mathcal{A}^*$ of length |W|. Every word W in $\mathcal{V}^{\#}$ has a <u>least shift</u>, $\mathcal{E}_{-1}(W)$, and a <u>greatest shift</u>, $\mathcal{E}_{+1}(W)$. That is $\mathcal{E}_{\pm 1}(W) = \sigma^k(W)$ with k taking the value in $\{0,...,|W|-1\}$ such that $O(\sigma^j(W), \sigma^k(W)) = \pm 1$ respectively for all j in $\{0,...,|W|-1\} \setminus \{k\}$. The least and greatest shifts are collectively called the <u>extreme shifts</u>. A word W in $\mathcal{V}^{\#}$ is maximal if $\mathcal{E}_{+1}(W) = W$, and minimal if $\mathcal{E}_{-1}(W) = W$. The term <u>$\Omega$-extremal</u> will mean minimal if $\Omega = -1$, and maximal if $\Omega = +1$. Note that W is $\Omega$-extremal if $O(\sigma^k(W), W) = \Omega$ for all k in $\{1,...,|W|-1\}$.

Finally we note that for any true statement about words in $\mathcal{V}^{\#}$ there exists another true statement, its <u>dual</u>, obtained from the former by <u>involution</u>, that is, by switching L and R, C and K, < and >, maximal and minimal, and the order of pairs. As an example, the $\chi$-seed (K,LC) would be replaced by the $\chi$-seed (RK,C).

## III.1 Proof of Theorem 1

### III.1.1 The Decide-or-Depute Lemma

Before presenting the pricipal tool we use in all the proofs, Lemma B, we make several elementary assertions whose proofs are simple. First we give a name to the fact that, in the comparison of a pair of words, what happens after the first discrepancy is of no consequence:

**Lemma A1:** If A, B, D and E are in $\mathcal{A}^* \cup \underline{\mathcal{A}}$, and $\tau_j(A) \neq \tau_j(B)$ for some j in $\{1,...,\min(|A|,|B|)\}$, then $O(A,B) = O(\tau_j(A), \tau_j(B)) = O(\tau_j(A)D, \tau_j(B)E)$.

Second, we note that when a terminal $Q \in \{C,K\}$ is changed to $Q^{\pm 1}$, the ordering and parities of the words produced are as given by the following lemma:

**Lemma A2:** For U in $\mathcal{B}^*$, $Q \in \{C,K\}$,
$$O(UQ^{-\rho(U)}..., UQ) = O(UQ, UQ^{+\rho(U)}...) = 1,$$
$$\rho(UQ^{\pm\rho(U)}) = \pm O(M,Q).$$

And third, we point out the following adjacency property:

**Lemma A3:** For U in $\mathcal{A}^*$, Q ∈ {C,K}, there is no word in $\mathcal{A}^*$ of length |UQ| between $UQ^{-1}$ and UQ, nor between UQ and $UQ^{+1}$.

We now introduce the decide-or-depute lemma: Lemma B. Theorem 1 asserts that certain products of the words {AP, $AP^{+1}$, $AP^{-1}$, BQ, $BQ^{+1}$, $BQ^{-1}$} (with AP and BQ in $\mathcal{V}^{\#}$, P and Q in {C,K}) are extremal as a consequence of the extremality and mutual bounding properties of the basic words AP and BQ. Proving it boils down to repeatedly determining the effect, on the order of a pair of words, of replacing a Q ∈ {C,K} in one of them by $Q^{\pm 1}$. Proving Lemmas D through G of section III.2 requires similar determinations. Lemma B, given below, tells us that when the Q is replaced by $Q^{\pm 1}$ either the order is unchanged, or it is determined in a specified way by the order of the trailing subwords beyond the position of the $Q^{\pm 1}$.

**Lemma B:** If UP and VQ are in $\mathcal{V}^{\#}$, P and Q are in {C,K}, and $\Omega \equiv \mathcal{O}(UP, VQ) \neq 0$, then for any D in $\mathcal{V}^{\#}$,

(i) $\mathcal{O}(UP^{-\Omega\rho(U)}D, VQ) = \Omega$, and
  either $\tau_{|UP|}(VQ) \neq UP^{+\Omega\rho(U)}$ and $\mathcal{O}(UP^{+\Omega\rho(U)}D, VQ) = \Omega$,
  or $\tau_{|UP|}(VQ) = UP^{+\Omega\rho(U)}$ and $\mathcal{O}(UP^{+\Omega\rho(U)}D, VQ) = \Omega\mathcal{O}(M,P)\mathcal{O}(D, \sigma^{|UP|}(VQ))$,

(ii) $\mathcal{O}(UP, VQ^{+\Omega\rho(U)}D) = \Omega$, and
  either $\tau_{|VQ|}(UP) \neq VQ^{-\Omega\rho(U)}$ and $\mathcal{O}(UP, VQ^{-\Omega\rho(U)}D) = \Omega$,
  or $\tau_{|VQ|}(UP) = VQ^{-\Omega\rho(U)}$ and $\mathcal{O}(UP, VQ^{-\Omega\rho(U)}D) = -\Omega\mathcal{O}(M,Q)\mathcal{O}(\sigma^{|VQ|}(UP), D)$.

**Proof:**
(i) Case |UP| > |VQ|.
  $\tau_{|UP|}(VQ) \neq UP$ because $|\tau_{|UP|}(VQ)| = |VQ| \neq |UP|$,
  and $\mathcal{O}(UP^{\pm}D, VQ) = \mathcal{O}(UP, VQ)$ by Lemma A1.
 Case |UP| = |VQ|.
  Case U=V.
   Then P≠Q, and and $\mathcal{O}(UP^{\pm}D, VQ) = \mathcal{O}(UP, VQ)$ by the observation that
   $\mathcal{O}(C,K) = \mathcal{O}(C^{\pm}, K) = \mathcal{O}(C, K^{\pm}) = 1$, and Lemma A1.
  Case U≠V.
   $\mathcal{O}(UP^{\pm}D, VQ) = \mathcal{O}(UP, VQ)$ by Lemma A1.
 Case |UP| < |VQ|.
  Case $\tau_{|U|}(V) \neq U$.

From Lemma A1, $\mathcal{O}(UP^{\pm}D,VQ) = \mathcal{O}(UP,VQ)$.

Case $\tau_{|U|}(V) = U$.

First note that $\mathcal{O}(UP,\tau_{|UP|}(VQ)) = \mathcal{O}(P,Q)$ by Lemma A1. Therefore
$$\Omega(\mathcal{N}(UP) + \Omega) \leq \Omega\mathcal{N}(\tau_{|UP|}(VQ)).$$
Now by Lemma A3,
$$\mathcal{N}(UP^{-\rho(U)}) + 1 = \mathcal{N}(UP) = \mathcal{N}(UP^{+\rho(U)}) - 1, \text{ whence}$$
$$\mathcal{N}(UP^{-\Omega\rho(U)}) + \Omega = \mathcal{N}(UP) = \mathcal{N}(UP^{+\Omega\rho(U)}) - \Omega.$$
Thus
$$\Omega\mathcal{N}(UP^{-\Omega\rho(U)}) < \Omega\mathcal{N}(\tau_{|UP|}(VQ)), \text{ and}$$
$$\Omega\mathcal{N}(UP^{+\Omega\rho(U)}) \leq \Omega\mathcal{N}(\tau_{|UP|}(VQ)).$$
Therefore $\Omega = \mathcal{O}(UP^{-\Omega\rho(U)},\tau_{|UP|}(VQ))$
$= \mathcal{O}(UP^{-\Omega\rho(U)}D,VQ)$ by Lemma A1.

If $<$ holds in the last inequality above, we are done, for then
$\tau_{|UP|}(VQ) \neq UP^{+\Omega\rho(U)}$, and
$\Omega = \mathcal{O}(UP^{+\Omega\rho(U)},\tau_{|UP|}(VQ))$
$= \mathcal{O}(UP^{+\Omega\rho(U)}D,VQ)$ by Lemma A1.

Otherwise (equality holds)
$\tau_{|UP|}(VQ) = UP^{+\Omega\rho(U)}$,
and since
$\rho(UP^{+\Omega\rho(U)}) = \Omega\mathcal{O}(M,P)$ by Lemma A2,
we have
$$\mathcal{O}(UP^{+\Omega\rho(U)}D,VQ) = \rho(UP^{+\Omega\rho(U)})\,\mathcal{O}(D,\sigma^{|UP|}(VQ))$$
$$= \Omega\mathcal{O}(M,P)\mathcal{O}(D,\sigma^{|UP|}(VQ)).$$

**(ii)** Switch UP and VQ in (i). ∎

### III.1.2.  <u>Common Comparison Lemmas</u>

As will be discussed in detail in section III.1.3, the extremality of the emanating words in Theorem 1 is established by comparing each emanating word with each of its shifts. These shifts fall into classes (for example: part of the way through the first A, beginning B, etc.), and each class is treated by the repeated application of the decide-or-depute lemma. To minimize repetition in the proof of Theorem 1, and in those of Lemmas D through G (were we to provide them), we note that several applications of Lemma B often lead to one of a small number of common comparisons, whose outcomes (themselves determined using Lemma B) are given here as Lemmas C. As mentioned earlier, the proofs are quite repetitious and we give only one of them (that of C3a) to serve as an example.

**Lemma C1:** If U and V are in $\mathcal{B}^*$, P and Q are in $\{C,K\}$, D is in $\mathcal{V}^\#$, and
(i) $\mathcal{O}(\sigma^j(UP),VQ) = -\mathcal{O}(M,P) \equiv \Omega$ for all $j \in \{0,...,|U|\}$,
(ii) $\mathcal{O}(\sigma^k(VQ),VQ) = \Omega$ for all $k \in \{1,...,|V|\}$,
then if $\tau_{|V|}(D) = V$,
$\mathcal{O}(\sigma^i(U)P^{\pm}D,VQ) = \Omega$ for all $i \in \{0,...,|U|\}$.

The following instances are perhaps more readable than the lemma itself:

If AC and BQ are in $\mathcal{V}^\#$, Q is in $\{C,K\}$, BQ is maximal, all shifts of AC are less than BQ, and begins B, then $\sigma^j(A)C^{\pm 1}D < BQ$ for all $j \in \{0,...,|A|\}$.

If AK and BQ are in $\mathcal{V}^\#$, Q is in $\{C,K\}$, BQ is minimal, all shifts of AK are greater than BQ, and D begins B, then $BQ < \sigma^j(A)K^{\pm 1}D$ for all $j \in \{0,...,|A|\}$.

**Lemma C2:** If VQ and D are in $\mathcal{V}^\#$, Q is in $\{C,K\}$, $\tau_{|V|}(D) = V$, and
$$\mathcal{O}(\sigma^i(V)Q,VQ) = -\mathcal{O}(M,Q) \equiv \Omega \text{ for all } i \in \{1,...,|V|\},$$
then
$$\mathcal{O}(\sigma^j(V)Q^{-\Omega\rho(V)}D,VQ) = \Omega \text{ for all } j \in \{0,...,|V|\},$$

Two instances are:
If AK and D are in $\mathcal{V}^\#$, AK is maximal and D begins A, then
$\sigma^j(A)K^{-\rho(A)}D < AK$ for all $j \in \{0,...,|A|\}$.

If BC and D are in $\mathcal{V}^\#$, BC is minimal and D begins B, then
$BC < \sigma^j(B)C^{+\rho(B)}D$ for all $j \in \{0,...,|B|\}$.

**Lemma C3:** **a.** If VQ in $\mathcal{V}^\#$ is $\Omega$-extremal, where $\Omega = -\mathcal{O}(M,Q)$, then
$$\mathcal{O}(\sigma^k(V)[Q^{-\Omega\rho(V)}V]^iQ, V[Q^{-\Omega\rho(B)}V]^jQ) = \Omega$$
for any $k \in \{1,...,|V|\}$, $i,j \in \{0,1,...\}$.

**b.** If VQ is in $\mathcal{V}^\#$, let $\Omega = -\mathcal{O}(M,Q)$, then
$$\mathcal{O}([VQ^{-\Omega\rho(V)}]^iQ, [VQ^{-\Omega\rho(V)}]^jQ) = \Omega$$
$i,j \in \{0,1,...\}$ iff $i > j$.

**Proof:** We arbitrarily choose Lemma C3a for an illustration of the methods of proof used for all the lemmas.
Let $\Omega(i,j,k) \equiv \mathcal{O}(\sigma^k(V)[Q^{-\Omega\rho(V)}V]^iQ, V[Q^{-\Omega\rho(B)}V]^jQ)$.
Then
$$\Omega(0,j,k) = \mathcal{O}(\sigma^k(V)Q, V[C^{-\rho(V)}V]^jQ)$$
$$= \mathcal{O}(\sigma^k(V)Q, \tau_{|\sigma^k(V)Q|}(V))$$

$$\begin{aligned} &= \mathcal{O}(\sigma^k(V)Q, VQ) \text{ by Lemma A1,} \\ &= \Omega \text{ by hypothesis.} \end{aligned}$$

$$\begin{aligned} \Omega(i,0,k) &= \mathcal{O}(\sigma^k(V)[Q^{-\rho(V)}V]^i Q, VQ) \\ &= \Omega \text{ by Lemma C2.} \end{aligned}$$

For $i > 0$ and $j > 0$, since by Lemma A1

$$\mathcal{O}(\sigma^k(V)Q, BD) = \Omega \text{ for any D in } \mathcal{V}^\#,$$

applying Lemma B yields,

$$\Omega(i,j,k) = \Omega \text{ if } \rho(\sigma^k(V)) = \rho(V),$$

and otherwise ($\rho(\sigma^k(V)) = -\rho(V)$)

$$\Omega(i,j,k) = \Omega$$

unless

$$\begin{aligned} \tau_{|\sigma^k(V)Q|}(V) &= \sigma^k(V)Q^{+\Omega\rho(\sigma^k(V))} \\ &= \sigma^k(V)Q^{-\Omega\rho(V)} \end{aligned}$$

in which case

$$\Omega(i,j,k) = \Omega \; \mathcal{O}(M,Q) \; \mathcal{O}(V[Q^{-\Omega\rho(V)}V]^{i-1}Q, \sigma^{k'}(V)[Q^{-\Omega\rho(B)}V]^j Q),$$

where $k' = |\sigma^k(V)Q|$,

$$\begin{aligned} &= -\Omega^2 \; (-\mathcal{O}(\sigma^{k'}(V)[Q^{-\Omega\rho(B)}V]^j Q, V[Q^{-\Omega\rho(V)}V]^{i-1}Q)) \\ &= \Omega(j, i-1, k'). \end{aligned}$$

By induction on i,j, $\Omega(i,j,k) = \Omega$.

∎

**Lemma C4:** If UQ and VQ are in $\mathcal{V}^\#$, with Q in {C,K}, and

$$\mathcal{O}(\sigma^i(VQ), VQ) = \mathcal{O}(\sigma^j(UQ), VQ) = -\mathcal{O}(M,Q) \equiv \Omega$$

for all $i \in \{1,...,|V|\}$, $j \in \{0,...,|U|\}$, then for all $n \in \{0,1,...\}$,

$$\mathcal{O}(\sigma^k(U)Q^{\pm}[VQ^{-\Omega\rho(V)}]^n VQ, [VQ^{-\Omega\rho(V)}]^m VQ) = \Omega$$

for any $k \in \{0,...,|U|\}$, $m \in \{0,1,...,n\}$.

### III.1.3 Proof of Theorem 1

We are now in a position to give a proof of the extremality of the emanating words of Theorem 1. The proof is long for three reasons: (i) there are altogether four different types of emanation, and for each type we must compare the emanation with all of its own shifts; (ii) the shifts of each emanation fall into a number of classes each of which must be treated separately; and (iii) each class of shifts must be treated with regard to the various possibilities concerning the relative length of compared subwords. However, the multiplicity of cases is really the only difficulty, and the proof is quite mechanical. Each case is handled by applying Lemma B repeatedly. The set of unresolved cases is systematically reduced, to consist of comparisons of progressively shorter pairs of words. Eventually the only unresolved comparison either is one of those given in the previous section, or involves only the basic words and is directly determined by the defining properties of the seed.

**Proof:**

    α.    Consider $[B]^nQ$, B in {L,R}, Q in {C,K}, n in {0,1,...}.
          If n=0, $[B]^nQ = Q$ is both minimal and maximal.
          If n>0, for j in {1,...,n}, $O(\sigma^j([B]^nQ),[B]^nQ) = O([B]^{n-j}Q,[B]^nQ)$
              $= \rho([B]^{n-j})O(Q,[B]^jQ) = 1 \cdot O(Q,B) = O(M,B)$.
          Therefore $[B]^nQ$ is $O(M,B)$-extremal.

    ψ.    Case $1 \leq k \leq |A|$. We must show that
          $O(AQ^{\pm\Omega\rho(A)}[BQ^{-\Omega\rho(b)}]^{n-1}BQ, \sigma^k(A)Q^{\pm\Omega\rho(A)}[BQ^{-\Omega\rho(B)}]^{n-1}BQ )$
                                               $= -O(M,Q) \equiv \Omega$.
          By Lemma A1 and hypothesis
            $O(AQ^{\pm\Omega\rho(A)}[BQ^{-\Omega\rho(b)}]^{n-1}BQ, \sigma^k(A)Q ) = \Omega$.
          So by Lemma B(ii),
            $O(AQ^{\pm\Omega\rho(A)}[BQ^{-\Omega\rho(b)}]^{n-1}BQ, \sigma^k(A)Q^{+\Omega\rho(\sigma^k(A))}[BQ^{-\Omega\rho(B)}]^{n-1}BQ ) = \Omega$,
          and
            $O(AQ^{\pm\Omega\rho(A)}[BQ^{-\Omega\rho(b)}]^{n-1}BQ, \sigma^k(A)Q^{-\Omega\rho(\sigma^k(A))}[BQ^{-\Omega\rho(B)}]^{n-1}BQ ) = \Omega$
          unless
            $\tau_{|\sigma^k(A)Q|}(AQ^{\pm\Omega\rho(A)}[BQ^{-\Omega\rho(b)}]^{n-1}BQ) = \sigma^k(A)Q^{-\Omega\rho(\sigma^k(A))}$
          in which case
            $O(AQ^{\pm\Omega\rho(A)}[BQ^{-\Omega\rho(b)}]^{n-1}BQ, \sigma^k(A)Q^{-\Omega\rho(\sigma^k(A))}[BQ^{-\Omega\rho(B)}]^{n-1}BC )$
            $= -\Omega \; O(M,Q) \; O(\sigma^{|\sigma^k(A)Q|}(A)Q^{\pm\Omega\rho(A)}[BQ^{-\Omega\rho(b)}]^{n-1}BQ, [BQ^{-\Omega\rho(b)}]^{n-1}BQ)$,
            $= (-\Omega)(-\Omega) \; \Omega$ by Lemma C4,
            $= \Omega$.

          Case $|A| < k$. We must show that for any k' in $\{0,...,|B|\}$, and $m \leq n$,
          $\Gamma \equiv O(AQ^{\pm\Omega\rho(A)}[BQ^{-\Omega\rho(b)}]^{n-1}BQ, \sigma^{k'}(B)[Q^{-\Omega\rho(B)}B]^{m-1}BQ ) = -O(M,Q) \equiv \Omega$.
          We have
            $O(AQ, \sigma^{k'}(B)Q ) = -O(M,Q) = \Omega$.
          by hypothesis.
          Case $|AQ| > |\sigma^{k'}(B)Q|$.
              $O(AQ^{\pm\Omega\rho(A)}[BQ^{-\Omega\rho(b)}]^{n-1}BQ, \sigma^{k'}(B)Q ) = \Omega$ by Lemma A1.
            Thus for m–1=0, $\Gamma = \Omega$.
            Otherwise (m–1 > 0), by Lemma B(i) if $\rho(\sigma^{k'}(B)) = -\rho(B)$, $\Gamma = \Omega$ directly.
            If $\rho(\sigma^{k'}(B)) = \rho(B)$,
                $\Gamma = O(AQ^{\pm\Omega\rho(A)}[BQ^{-\Omega\rho(b)}]^{n-1}BQ, \sigma^{k'}(B)[Q^{-\Omega\rho(B)}B]^{m-1}Q )$

$$= \mathcal{O}(AQ^{\pm\Omega\rho(A)}[BQ^{-\Omega\rho(b)}]^{n-1}BQ, \sigma^{k'}(B)[Q^{-\Omega\rho(\sigma^{k'}(B))}B]^{m-1}Q)$$
$$= \Omega$$

unless
$$\tau_{|\sigma^{k'}(B)Q|}(AQ^{\pm\Omega\rho(A)}[BQ^{-\Omega\rho(b)}]^{n-1}BQ) = \sigma^{k'}(B)Q^{-\Omega\rho(\sigma^{k'}(B))}$$
in which case
$$\Gamma =$$
$$-\Omega \; \mathcal{O}(M,Q) \; \mathcal{O}(\sigma^{|\sigma^{k'}(B)Q|}(AQ^{\pm\Omega\rho(A)}[BQ^{-\Omega\rho(b)}]^{n-1}BQ), B[Q^{-\Omega\rho(B)}B]^{m-1}Q)$$
$$= (-\Omega)(-\Omega)\,\Omega \text{ by Lemma C3a,}$$
$$= \Omega.$$

Case $|AQ| = |\sigma^{k'}(B)Q|$.

The hypotheses imply that $\mathcal{O}(A, \sigma^{k'}(B)) = \Omega$. So $\Gamma = \Omega$ by Lemma A1.

Case $|AQ| < |\sigma^{k'}(B)Q|$.

We have $\Omega = \mathcal{O}(AQ, \sigma^{k'}(B)Q)$
$$= \mathcal{O}(AQ, \sigma^{k'}(B)[Q^{-\Omega\rho(B)}B]^{m-1}Q).$$

So by Lemma B(i),
$$\mathcal{O}(AQ^{-\Omega\rho(A)}..., \sigma^{k'}(B)[Q^{-\Omega\rho(B)}B]^{m-1}Q) = \Omega,$$
and
$$\mathcal{O}(AQ^{+\Omega\rho(A)}..., \sigma^{k'}(B)[Q^{-\Omega\rho(B)}B]^{m-1}Q) = \Omega$$
unless
$$\tau_{|AQ|}(\sigma^{k'}(B)) = AQ^{+\Omega\rho(A)},$$
in which case
$$\mathcal{O}(AQ^{+\Omega\rho(A)}..., \sigma^{k'}(B)[Q^{-\Omega\rho(B)}B]^{m-1}Q)$$
$$= \quad \Omega \; \mathcal{O}(M,Q) \; \mathcal{O}(BQ^{-\Omega\rho(b)}]^{n-1}BQ, \sigma^{k'+|AQ|}(B)[Q^{-\Omega\rho(B)}B]^{m-1}Q)$$
$$= \quad \Omega \; (-\Omega)(-\Omega) \text{ by Lemma C3a,}$$
$$= \quad \Omega.$$

χ.     Let (AK,BC) be a χ-seed.

   **(i)**    We show that $\beta_{(+1,\pm1)}(AK,BC)$ are maximal.

That is, we show $\mathcal{O}(\sigma^k(AK^{\pm\rho(A)}BC), AK^{\pm\rho(A)}BC) = 1$, for any k in $\{1,...,|AK^{\pm}B|\}$.

Case $1 \leq k \leq |A|$:
$$\mathcal{O}(\sigma^k(AK^{\pm\rho(A)}BC), AK^{\pm\rho(A)}BC) = \mathcal{O}(\sigma^k(A)K^{\pm\rho(A)}BC, AK^{\pm\rho(A)}BC).$$
Since $\mathcal{O}(\sigma^k(A)K, AK^{\pm\rho(A)}BC) = 1$, (by hypothesis because $|\sigma^k(A)| < |A|$),
by Lemma B(i) we have
$$\mathcal{O}(\sigma^k(A)K^{-\rho(\sigma^k(A))}BC, AK^{\pm\rho(A)}BC) = 1,$$
and
$$\mathcal{O}(\sigma^k(A)K^{+\rho(\sigma^k(A))}BC, AK^{\pm\rho(A)}BC,) = 1$$
unless
$$\tau_{|\sigma^k(A)K|}(AK^{\pm\rho(A)}BC) = \sigma^k(A)K^{+\rho(\sigma^k(A))}$$
in which case

$$O(\sigma^k(A)K^{+\rho(\sigma^k(A))}BC, AK^{\pm\rho(A)}BC)$$
$$= +1 \cdot O(M,K) O(BC, \sigma^{|\sigma^k(A)K|}(AK^{\pm\rho(A)}BC))$$
$$= O(BC, \sigma^{|\sigma^k(A)K|}(A)K^{\pm\rho(A)}BC)$$
$$= 1, \text{ by Lemma C1.}$$

Case $k = |AK|$.  $O(BC, AK^{\pm\rho(A)}BC) = 1$, by Lemma C1.

Case $|AK| < k$. We must show $O(\sigma^{k'}(BC), AK^{\pm}BC) = 1$, $k'$ in $\{0,...,|B|\}$.

Since $O(\sigma^{k'}(BC), AK) = 1$, then by Lemma B(ii)
$$O(\sigma^{k'}(BC), AK^{+\rho(A)}BC) = 1,$$
and
$$O(\sigma^{k'}(BC), AK^{-\rho(A)}BC) = 1,$$
unless
$$\tau_{|AK|}(\sigma^{k'}(BC)) = AK^{-\rho(A)}$$
in which case
$$O(\sigma^{k'}(BC), AK^{-\rho(A)}BC) = -1 \cdot O(M,K) O(\sigma^{|AK|}(\sigma^{k'}(BC)), BC) = -(-1) = 1.$$

That $\beta_{(-1,\pm 1)}(AK, BC)$ are minimal is the dual of the statement just proved.

**(ii)** We show that for all $n$ in $\{1,2,...\}$, $\chi_{(-1,\pm n)}(AK, BC)$ are minimal. The following discussion applies to any fixed $n$. Setting $VC \equiv AK^{-\pm\rho(A)}BC$, we must show that
$$\Delta_j \equiv O(BC^{-\rho(B)}[VC^{-\rho(V)}]^{n-1}VC, \sigma^j(BC^{-\rho(B)}[VC^{-\rho(V)}]^{n-1}VC)) = 1$$
for all $j$ in $\{1,...,|BC^{-\rho(B)}[VC^{-\rho(V)}]^{n-1}V|\}$.

Case $1 \le j \le |B|$.

Then $\Delta_j = O(BC^{-\rho(B)}[VC^{-\rho(V)}]^{n-1}VC, \sigma^j(B)C^{-\rho(B)}[VC^{-\rho(V)}]^{n-1}VC)$.

We have $O(BC^{-\rho(B)}, \sigma^j(B)C) = 1$, by hypothesis and Lemma A1,

so by Lemma B(ii)

$\Delta_j = 1$ directly or ($\rho(\sigma^j(B)) = \rho(B)$ and)
$$\Delta_j = -1 \cdot O(M,C) \cdot O(\sigma^{|\sigma^j(B)C|}(B)C^{-\rho(B)}[VC^{-\rho(V)}]^{n-1}VC, [VC^{-\rho(V)}]^{n-1}VC),$$
$$= -1 \cdot O(M,C) \cdot 1, \text{ by Lemma C4 and Theorem } 1\chi(i).$$
$$= 1.$$

Case $j > |B|$. Then either

$\Delta_j = O(BC^{-\rho(B)}[VC^{-\rho(V)}]^{n-1}VC, \sigma^{j'}(V)C) = 1$ by hypothesis, or

$\Delta_j = \Delta'_{j'} \equiv O(BC^{-\rho(B)}[VC^{-\rho(V)}]^{n-1}VC, \sigma^{j'}(V)C^{-\rho(V)}[VC^{-\rho(V)}]^{m-1}VC)$

for some $j'$ in $\{0,...,|V|\}$ and $m$ in $\{1,...,n-1\}$.

Case $j' = |AK|$.  $\Delta = O(BC^{-\rho(B)}[VC^{-\rho(V)}]^{n-1}VC, BC^{-\rho(V)}[VC^{-\rho(V)}]^{m-1}VC)$.

Case $\rho(B) = -\rho(V)$.

$\Delta'_{j'} = 1$ since $O(BC^{-\rho(B)}, BC^{+\rho(B)}) = 1$.

Case $\rho(B) = \rho(V)$.

$\Delta'_{j'} = O(BC^{-\rho(B)}) O([VC^{-\rho(V)}]^{n-1}VC, [VC^{-\rho(V)}]^{m-1}VC)$,
$$= O(BC^{-\rho(B)}) \cdot 1, \text{ by Lemma C3b,}$$
$$= 1.$$

Case $j' > |AK|$. Then

$\Delta_{j'}' = \Delta_{j''}'' \equiv \mathcal{O}(BC^{-\rho(B)}[VC^{-\rho(V)}]^{n-1}VC, \sigma^{j''}(B)C^{-\rho(V)}[VC^{-\rho(V)}]^{m-1}VC)$,

for some j'' in $\{1,...,|B|\}$. We have by hypothesis and Lemma A1

$\mathcal{O}(BC^{-\rho(B)}[VC^{-\rho(V)}]^{n-1}VC, \sigma^{j''}(B)C) = 1$.

Case $\rho(B) = -\rho(V)$.

$\Delta_{j''}'' = 1$ by Lemma B(ii).

Case $\rho(B) = \rho(V)$.

By Lemma B(ii) $\Delta_{j''}'' = 1$ directly, or

$\Delta_{j''}'' = \mathcal{O}(\sigma^{|\sigma^{j''}(B)C|}(B)C^{-\rho(B)}[VC^{-\rho(V)}]^{n-1}VC, [VC^{-\rho(V)}]^{m-1}VC)$

$= 1$ by Lemma C4.

Case $0 \leq j' \leq |A|$. Then

$\Delta_{j'}' = \mathcal{O}(BC^{-\rho(B)}[VC^{-\rho(V)}]^{n-1}VC, \sigma^{j'}(A)K^{\pm\rho(A)}BC^{\pm\rho(B)}[VC^{-\rho(V)}]^{m-1}VC)$.

Case $|\sigma^{j'}(A)| < |B|$. We have by hypothesis and Lemma A1

$\mathcal{O}(BC^{-\rho(B)}[VC^{-\rho(V)}]^{n-1}VC, \sigma^{j'}(A)K) = 1$

Therefore by Lemma B,

$\mathcal{O}(BC^{-\rho(B)}[VC^{-\rho(V)}]^{n-1}VC, \sigma^{j'}(A)K^{+\rho(\sigma^{j'}(A))}BC^{\pm\rho(B)}[VC^{-\rho(V)}]^{m-1}VC) = 1$,

and

$\mathcal{O}(BC^{-\rho(B)}[VC^{-\rho(V)}]^{n-1}VC, \sigma^{j'}(A)K^{-\rho(\sigma^{j'}(A))}BC^{\pm\rho(B)}[VC^{-\rho(V)}]^{m-1}VC) = 1$

directly or

$= -1.\mathcal{O}(M,K).\mathcal{O}(\sigma^{j'''}(B)C^{-\rho(B)}[VC^{-\rho(V)}]^{n-1}VC, BC^{\pm\rho(B)}[VC^{-\rho(V)}]^{m-1}VC)$

$= \mathcal{O}(BC^{\pm\rho(B)}[VC^{-\rho(V)}]^{m-1}VC, \sigma^{j'''}(B)C^{-\rho(B)}[VC^{-\rho(V)}]^{n-1}VC)$,

for some j''' in $\{1,...,|\sigma^{j'}(A)K| \leq |B|\}$.

Now we have

$\mathcal{O}(BC^{\pm\rho(B)}[VC^{-\rho(V)}]^{m-1}VC, \sigma^{j'''}(B)C) = 1$ by hypothesis and Lemma A1,

so by Lemma B

$\mathcal{O}(BC^{\pm\rho(B)}[VC^{-\rho(V)}]^{m-1}VC, \sigma^{j'''}(B)C^{+\rho(\sigma^{j'''}(B))}[VC^{-\rho(V)}]^{n-1}VC) = 1$,

and

$\mathcal{O}(BC^{\pm\rho(B)}[VC^{-\rho(V)}]^{m-1}VC, \sigma^{j'''}(B)C^{-\rho(\sigma^{j'''}(B))}[VC^{-\rho(V)}]^{n-1}VC) = 1$,

directly or

$= -1.\mathcal{O}(M,C).\mathcal{O}(\sigma^{j'''}(B)C^{\pm\rho(B)}[VC^{-\rho(V)}]^{m-1}VC, [VC^{-\rho(V)}]^{n-1}VC)$,

$= \mathcal{O}(\sigma^{j'''}(B)C^{\pm\rho(B)}[VC^{-\rho(V)}]^{m-1}VC, [VC^{-\rho(V)}]^{n-1}VC)$.

Now we have $\mathcal{O}(\sigma^{j'''}(B)C, [VC^{-\rho(V)}]^{n-1}VC) = 1$ by hypothesis and Lemma A1, so

$\mathcal{O}(\sigma^{j'''}(B)C^{-\rho(\sigma^{j'''}(B))}[VC^{-\rho(V)}]^{m-1}VC, [VC^{-\rho(V)}]^{n-1}VC) = 1$,

and

$\mathcal{O}(\sigma^{j'''}(B)C^{+\rho(\sigma^{j'''}(B))}[VC^{-\rho(V)}]^{m-1}VC, [VC^{-\rho(V)}]^{n-1}VC) = 1$

directly or

$= +1.\mathcal{O}(M,C).\mathcal{O}([VC^{-\rho(V)}]^{m-1}VC, \sigma^{j''''}(V)C^{-\rho(V)}[VC^{-\rho(V)}]^{n'-1}VC)$,

for some $n' \geq m$, and where $j'''' = |\sigma^{j'''}(B)C|$,

$= -1.-1$ by Lemma C3a,

$= 1$.

Case $|\sigma^{j'}(A)| = |B|$. We have
$\mathcal{O}(BC,\sigma^{j'}(A)K) = 1$ by hypothesis, and so
$\mathcal{O}(BC^{-\rho(B)}[VC^{-\rho(V)}]^{n-1}VC,\sigma^{j'}(A)K) = 1$
Then by Lemma B(ii)
$\mathcal{O}(BC^{-\rho(B)}[VC^{-\rho(V)}]^{n-1}VC,\sigma^{j'}(A)K^{+\rho(\sigma^{j'}(A))}BC^{\pm\rho(B)}[VC^{-\rho(V)}]^{m-1}VC)$
$= 1$, and
$\mathcal{O}(BC^{-\rho(B)}[VC^{-\rho(V)}]^{n-1}VC,\sigma^{j'}(A)K^{-\rho(\sigma^{j'''}(B))}BC^{\pm\rho(B)}[VC^{-\rho(V)}]^{m-1}VC)$
$= 1$ directly or
$= -\mathcal{O}(M,K)\,\mathcal{O}([VC^{-\rho(V)}]^{n-1}VC,BC^{\pm\rho(B)}[VC^{-\rho(V)}]^{m-1}VC)$
$= \mathcal{O}(BC^{\pm\rho(B)}[VC^{-\rho(V)}]^{m-1}VC,[VC^{-\rho(V)}]^{n-1}VC)$
Now $\mathcal{O}(BC,[VC^{-\rho(V)}]^{n-1}VC) = 1$ by hypothesis and Lemma A1, so
$\mathcal{O}(BC^{-\rho(B)}[VC^{-\rho(V)}]^{m-1}VC,[VC^{-\rho(V)}]^{n-1}VC) = 1$, and
$\mathcal{O}(BC^{+\rho(B)}[VC^{-\rho(V)}]^{m-1}VC,[VC^{-\rho(V)}]^{n-1}VC) = 1$
unless $BC^{+\rho(B)} = \tau^{|BC|}(V)$, in which case
$\mathcal{O}(BC^{+\rho(B)}[VC^{-\rho(V)}]^{m-1}VC,[VC^{-\rho(V)}]^{n-1}VC)$
$= \mathcal{O}(M,C)\,\mathcal{O}([VC^{-\rho(V)}]^{m-1}VC,\sigma^{|BC|}(V)C^{-\rho(V)}[VC^{-\rho(V)}]^{n-2}VC)$ if $n>1$,
$= -1.-1 = 1$ by Lemma C3a,
or, if $n=1$,
$= \mathcal{O}(M,C)\,\mathcal{O}([VC^{-\rho(V)}]^{m-1}VC,\sigma^{|BC|}(V)C) = 1$ by hypothesis.
Case $|\sigma^{j'}(A)| > |B|$. We have by hypothesis and Lemma A1
$\mathcal{O}(BC,\sigma^{j'}(A)K^{-\pm\rho(A)}BC^{\pm\rho(B)}[VC^{-\rho(V)}]^{m-1}VC)$,
so by Lemma B
$\Delta = \mathcal{O}(BC^{-\rho(B)}[VC^{-\rho(V)}]^{n-1}VC,\sigma^{j'}(A)K^{-\pm\rho(A)}BC^{\pm\rho(B)}[VC^{-\rho(V)}]^{m-1}VC)$
$= 1$.
That $\chi_{(+1,\pm n)}(AK,BC)$ are maximal is the dual of the statement just proven.

∎

## III.2   Proof of Theorem 2

### III.2.1  Extreme-shift Lemmas

The proof of Theorem 2, to be given shortly, is constructive. It uses a deterministic algorithm (the Retraction Algorithm; see III.2.4) to effect a decomposition of an extremal word in $\mathcal{V}^{\#}$ that reveals its seed of origin. The "retraction" to the seed entails the repeated truncation of the word—deletion of the extreme shift—and it relies both on the extremality of the word remaining at each stage of the process, and on knowledge of the extreme shift of the remaining word. The first lemma in this section identifies the extreme shifts of the extremal words that are specified as emanations in Theorem 1. Those that follow assure the extremality of the words that remain after each truncation, and enable us to identify the extreme shifts of those words.

**Lemma D:**   α.   For each n in $\{1,2,...\}$, $\mathcal{E}_{+1}([L]^{n-1}C) = C$ and $\mathcal{E}_{-1}([R]^{n-1}K) = K$.

ψ.   If $(UQ, VQ)$ is a ψ-seed, then with $\Omega = -\mathcal{O}(M,Q)$, for each n in $\{1,2,...\}$, $\mathcal{E}_{\Omega}(\psi_{\pm n}(UQ,VQ)) = VQ$.

χ.   If $(AK,BC)$ is a χ-seed, then for each n in $\{1,2,...\}$,
the greatest shift of $\beta_{(-1,\pm 1)}(AK,BC)$ is AK,
the greatest shift of $\chi_{(-1,\pm n)}(AK,BC)$ is $AK^{-(\pm\rho(A))}BC$,
the least shift of $\beta_{(+1,\pm 1)}(AK,BC)$ is BC, and
the least shift of $\chi_{(+1,\pm n)}(AK,BC)$ is $BC^{\pm\rho(B)}AK$.

**Lemma E1:**   a.   If UXVC, with $X \in \mathcal{B}$, is in $\mathcal{V}^{\#}$ and is minimal, and WC is its greatest shift, then UC is minimal.

b.   If UXVC, with $X \in \mathcal{B}$, is in $\mathcal{V}^{\#}$ and maximal, and VC is its least shift, then UK is maximal.

**Lemma E2:**   If UVXVC, with $X \in \mathcal{B}$, is in $\mathcal{V}^{\#}$ and minimal with greatest shift VC, then UVC is minimal with greatest shift VC.

**Lemma E3:**   If UYVXUC, with $X,Y \in \mathcal{B}$, is in $\mathcal{V}^{\#}$ and minimal with greatest shift VXUC, then UC is the least shift of VXUC.

### III.2.2 Uniqueness lemma

The Retraction Algorithm (III.2.4) ignores the letter that precedes the extreme shift at each stage of the decomposition, except the last. The following lemma is needed to show that knowledge of those letters is nevertheless not lost. The reason is that the extremal words specified in Theorem 1 are unique in the following senses.

**Lemma F:**    $\alpha$.    $[L]^n C$, is the only word in $\mathcal{V}^{\#}$ of length $|[L]^n C|$ with greatest shift C, $n \in \{0,1,...\}$, and $[R]^n K$, is the only word in $\mathcal{V}^{\#}$ of length $|[R]^n K|$ with least shift K.

        $\psi$.    If $(AQ, BQ)$ is a $\psi$-seed, the words $\Psi_n^{\pm}(AQ,BQ)$ defined in Theorem 1 are the only words in $\mathcal{V}^{\#}$ of the form
$$AX_0 BX_1 BX_2 ... BX_n BQ, \quad \text{with } X_i \in \mathcal{B}$$
for which $\mathcal{E}_{-O(M,Q)}(AX_0 BX_1 BX_2 ... BX_n BQ) = BQ$.

        $\chi$.    If $(AK, BC)$ is a $\chi$-seed, then
$$\chi_{(-1,\pm n)}(AK, BC), \ n \in \{1,2,...\},$$
are the only minimal words in $\mathcal{V}^{\#}$ of the form
$$W^{\pm}_{\mathbf{x}} = BX_0 \ AK^{\pm}BX_1 \ AK^{\pm}BX_2 \ ... AK^{\pm}BX_{n-1} \ AK^{\pm}BC$$
for which $\mathcal{E}_{+1}(W^{\pm}_{\mathbf{x}}) = AK^{+\rho(A)}BC$, and
$$\chi_{(+1,\pm n)}(AK, BC), \ n \in \{1,2,...\},$$
are the only maximal words in $\mathcal{V}^{\#}$ of the form
$$W^{\pm}_{\mathbf{x}} = AX_0 \ BC^{\pm}AX_1 \ BC^{\pm}AX_2 \ ... BC^{\pm}AX_{n-1} \ BC^{\pm}AK$$
for which $\mathcal{E}_{-1}(W^{\pm}_{\mathbf{x}}) = BC^{-\rho(B)}AK$.

### III.2.3 Seed existence lemmas

The Retraction Algorithm eventually generates a pair of words in $\mathcal{V}^{\#}$, $(UP, VQ)$, with $P, Q \in \{C, K\}$, which is asserted to be the seed from which the original word emanates. The following lemmas are required to ensure that the pair generated does in fact have the defining properties of a seed of the appropriate type.

**Lemma G1:**    **a.**    If $AXBC$ in $\mathcal{V}^{\#}$, with $X \in \mathcal{B}$, has least shift BC, then $O(BC, \sigma^j(AK)) = 1$ for all $j \in \{0,...,|A|\}$.

  **b.** If AXBC in $\mathcal{V}^\#$, with X$\in \mathcal{B}$, is maximal with least shift BC, then
    $\mathcal{O}(\sigma^j(BC),AK) = 1$ for all $j \in \{0,...,|B|\}$.

**Lemma G2:** **a.** If AXBC is in $\mathcal{V}^\#$, with X$\in \mathcal{B}$, is minimal with greatest shift BC, and A is not a shift of B, then $\mathcal{O}(AC,\sigma^j(BC)) = 1$ for all $j \in \{0,...,|B|\}$.

  **b.** If AXBC is in $\mathcal{V}^\#$, with X$\in \mathcal{B}$, is minimal with greatest shift BC, and B is not a shift of A, then $\mathcal{O}(\sigma^j(AC),BC) = 1$ for all $j \in \{0,...,|A|\}$.

**Lemma H1:** If AXBC in $\mathcal{V}^\#$, with X$\in \mathcal{A}'$, is maximal with least shift BC, then (AK,BC) is a $\chi$-seed.
**Proof:** BC is minimal by hypothesis.
  AK is maximal by Lemma E1b.
  $\mathcal{O}(BC,\sigma^j(AK)) = 1$ for all $j \in \{0,...,|A|\}$ by Lemma G1a.
  $\mathcal{O}(\sigma^j(BC),AK) = 1$ for all $j \in \{0,...,|B|\}$ by Lemma G1b.

**Lemma H2:** If AXBC in $\mathcal{V}^\#$, with X$\in \mathcal{B}$, is minimal with greatest shift BC, and neither A nor B is a shift of the other, then (BC,AC) is a $\psi$-seed.
**Proof:** BC is maximal by hypothesis.
  AC is minimal by Lemma E1a.
  $\mathcal{O}(\sigma^j(AC),BC) = 1$ for all $j \in \{0,...,|A|\}$ by Lemma G2b.
  $\mathcal{O}(AC,\sigma^j(BC)) = 1$ for all $j \in \{0,...,|B|\}$ by Lemma G2a.

### II.2.4 Proof of Theorem 2

We are finally in a position to present a proof of our result that every word in $\mathcal{V}^\#$ emanates from a unique seed. Given a word in $\mathcal{V}^\#$, the algorithm given below determines a seed from which the word emanates, as well as the complete specification of the emanation. The letters $X_1$, $X_2,...,X_n$ of Lemma F are ignored by the algorithm, but by that lemma, those letters are determined by $X_0$, which is not ignored. Existence of a seed of origin is thus constructively demonstrated. Using Lemma D, it is easily checked that all the emanations specified in Theorem 1 are correctly decomposed by the algorithm. That is, the A and B words are recovered, the type of seed is identified, and the type and indices of the emanation are determined correctly. Uniqueness therefore follows from the fact that if a word emanates from the seed (AP,BQ) and from the seed (A'P',B'Q'), then the algorithm recovers (AP,BQ) and it recovers (A'P',B'Q'). Since the algorithm is deterministic, (AP,BQ)=(A'P',B'Q').

**Retraction Algorithm:**

The seed from which a word in $\mathcal{V}^{\#}$ emanates is determined by applying the following algorithm.

Case: The word is of the form WC.

    Case: WC is maximal.

        Case: $|W|=0$.

            WC = C is also minimal, and WC = $\alpha_{-1}$.

        Case: $|W|\neq 0$.

            Define U,V in $\mathcal{B}^*$, X in $\mathcal{B}$, by WC=UXVC, where VC is the least shift of WC.

            Then by Lemma H1, (UK,VC) is a $\chi$-seed, and

$$WC = \beta_{(+1,\rho(UX))}(UK,VC).$$

    Case: WC is minimal.

        I.    If the greatest shift of WC is C, then WC= $\alpha_{-|WC|}$, by Lemma F$\alpha$.

            Otherwise set TC = WC, and VC = $E_{+1}$(WC) which is maximal.

        II.   If VC is not a shift of TC go to step III.

            Otherwise, define UX, X in $\mathcal{B}$, by TC = UXVC. By Lemma E1, UC is minimal.

            By Lemma E2, if VC is a shift of UC, it is its greatest shift. Redefine T by

            TC=UC and repeat step II.

        III.  Let n be the number of times the body of step II was performed.

            If TC is a shift of VC, it is its least shift by Lemma E3.

                Therefore defining SX, with X in $\mathcal{B}$, by VC = SXTC, (SK,TC) is a

                $\chi$-seed by Lemma H1, and WC= $\chi_{(-1,-\rho(SX)n)}$(SK,TC).

            Otherwise, neither TC nor VC is a shift of the other, whence by Lemma H2,

            (VC,TC) is a $\psi$-seed, and WC = $\Psi_{-\rho(TX)n}$(VC,TC).

    (Note: TC cannot become empty by removal of VC because TC$\neq$VC since one is maximal and the other is minimal, and VC is not the one word, C, that is both. That possibility was disposed of in step I.)

Case: The word is of the form WK.

    The seed is determined by the dual under involution of the algorithm above.

**Examples:**

(i) We apply the algorithm to the word LLMMMC, which is minimal.

    I.      Set TC = LLMMMC, and VC = MC the greatest shift.

    II.     VC is a shift of TC, so set UX = LLMM, and redefine TC = UC = LLMC.

            VC is a shift of TC, so set UX = LL, and redefine TC = UC = LC.

            VC is not a shift of TC, so proceed to step III.

    III.    Now VC = MC, TC = LC, X=L, n=2, and $\rho(TX) = \rho(LL) = +1$.

            TC is not a shift of VC, therefore LLMMMC = $\Psi_{-2}$(MC,LC).

(ii) We apply the algorithm to the word LMC, which is minimal.

    I.      Set TC = LMC, and VC = MC the greatest shift.

    II.     VC is a shift of TC, so set UX = L, and redefine TC = UC = C.

            VC=MC is not a shift of TC=C, so proceed to step III.

    III.    Now VC = MC, TC = C, X=L, n=1, and $\rho(TX) = \rho(L) = +1$.

            TC is a shift of VC, so set SX by MC = SXTC, that is S=1 and X=M,

            and so LMC = $\chi_{(-1,-1)}$(K,C).

                                                                                                                                     ■

## IV. A new partial order on finite kneading sequences

Theorems 1 and 2 yield directly a new partial order on finite kneading sequences that we formalize as follows:

- say $K_1$ is a <u>strict parent</u> of $K_2$ if $K_2$ emanates from a seed of the form $(K_0,K_1)$ or $(K_1,K_0)$, (by Theorem 2, each finite kneading sequences has exactly two strict parents),

- say $K_1$ is a <u>parent</u> of $K_2$, in symbols $K_1 \ll K_2$, if $K_1$ is a strict parent of $K_2$, or $K_1 = K_2$.

- say $K_1$ is an <u>ancestor</u> of $K_n$, in symbols $K_1 \ll\ll K_n$, if there exists a sequence $(K_1,K_2,...K_n)$, with $n \geq 2$, such that $K_i \ll K_{i+1}$ for $1 \leq i \leq n-1$.

Defining the <u>topological entropy h(K) of a kneading sequence K</u> as the infimum of topological entropies of maps f such that K is a kneading sequence of f, we have:

**Theorem 3:**  $K_1 \ll\ll K_2 \Rightarrow h(K_1) \leq h(K_2)$.

The two-parameter family of stunted saw-tooth maps are defined in the next section. Since the kneading data of a map determines its topological entropy[MT], and since any kneading data for $\{+-+\}$-bimodal maps is realized by some stunted sawtooth map, we can also define the topological entropy $h(K)$ of a kneading sequence K as the infimum of topological entropies of stunted saw-tooth maps $T^+_{a,b}$ such that K is a kneading sequence of $T^+_{a,b}$. The proof of Theorem 3 will be given in the next section, using this alternate but equivalent definition of $h(K)$.

## V. Stunted saw-tooth maps

We set define the <u>full $\{+-+\}$-bimodal stunted saw-tooth map</u> $S^+$ as:

$$S^+ = \begin{cases} 5x & \text{if } 0 \leq x \leq \frac{1}{5}, \\ 5(\frac{3}{5} - x) & \text{if } \frac{2}{5} \leq x \leq \frac{3}{5}, \\ 5(x - \frac{4}{5}) & \text{if } \frac{4}{5} \leq x \leq 1, \\ 1 & \text{if } \frac{1}{5} \leq x \leq \frac{2}{5}, \\ 0 & \text{if } \frac{3}{5} \leq x \leq \frac{4}{5}. \end{cases}$$

(The symbol $S^-$ will be reserved for the analogous $\{-+-\}$-bimodal map, treated elsewhere[R].) The graph of $S^+$ is a succession of a hill and a valley. It is easy to check that the set of itineraries for this map is the set of all possible itineraries for $\{+-+\}$-bimodal maps. We can now define a two-parameter family of $\{+-+\}$-bimodal stunted saw-tooth maps $T^+_{a,b}$ by setting

$$T^+_{a,b}(x) = \begin{cases} \min(S^+(x), a) & \text{under the hill,} \\ \max(S^+(x), 1-b) & \text{in the valley.} \end{cases}$$

For the map $T^+_{a,b}$ to be an endomorphism of the unit interval imposes the constraints

$$0 \leq a \leq 1, 0 \leq b \leq 1, \ a+b \geq 1. \tag{*}$$

The set of parameter values (a,b) satisfying (*) is denoted by $\mathbf{P}^+$. Figure 1.0 shows the graph of $S^+$ and the parameter space $\mathbf{P}^+$. All kneading sequences of the truncated maps $T^+_{a,b}$ can be understood as the itineraries under $S^+$ of those points which are closer to one of the turning points than any other point in their orbit. This makes the study of these maps quite simple: the arguments presented in [GT] and [BMT] for stunted maps in the unimodal case extend easily; in particular, any {+-+} kneading data occurs for some pair (a,b). More on general stunted families will be reported in [DGMT]; here we use these maps for the sake of illustration and to provide a geometric picture to display the results of Theorems 2 and 3.

In $\mathbf{P}^+$, following [MaT1-MaT3], [M] and [RS1, RS2] we define subsets corresponding to finite kneading sequences, which can be described as follows (proofs are straightforward and left to the reader):

- Any finite maximal kneading sequence terminating K is realised on a single segment of straight line parallel to the a-axis, with one <u>external</u> end point on the line b=1, and one end point called <u>internal</u> in the interior of $\mathbf{P}^+$ or on the boundary a+b=1: this segment is called a <u>K-ligament</u>.

- Any finite minimal kneading sequence terminating C is realised on a single segment of straight line parallel to the b-axis, with one <u>external</u> end point on the line a=1, and one end point called <u>internal</u> in the interior of $\mathbf{P}^+$ or on the boundary a+b=1: this segment is callet a <u>C-ligament</u>; we say <u>ligament</u> for either K-ligament or C-ligament.

- For any n>1 and a permutation **s** on n elements obtainable by restricting a {+-+} bimodal map to one of its periodic orbits, the <u>bone</u> $B_{\mathbf{s}}$ is the set of pairs (a,b) such that for $T^+_{a,b}$, one of the turning points belongs to a periodic orbit where the restriction of $T^+_{a,b}$ acts like **s**. Any bone is made of two connected curves $B^C_{\mathbf{s}}$ and $B^K_{\mathbf{s}}$ where the superscript indicates which turning point belongs to the periodic orbit. The curves $B^C_{\mathbf{s}}$ and $B^K_{\mathbf{s}}$ intersect at two points, one of which is a $\chi$-point, the other one corresponding to the coexistence of two periodic orbits with the same **s**, each of which contains a turning point. The curve $B^C_{\mathbf{s}}$ has two segments parallel to the a-axis, each havingwith one <u>external</u> end point on the line b=1. The interior end points of these segments are joined by a piece of K-ligament which belongs to $B^C_{\mathbf{s}}$. The curve $B^K_{\mathbf{s}}$ has two segments parallel to the b-axis, each havingwith one <u>external</u> end point on the line a=1. The interior end points of these segments are joined by a piece of C-ligament which belongs to $B_s^K$.

Figures 1.1, 1.2 and 1.3 represent (not to scale) respectively $\alpha$, $\chi$ and $\psi$ points, the ligaments which cross at these points, and the ligaments and pieces of bones so generated in **P**$^+$. These figures give a geometrical interpretation of Lemma D and of Theorems 1 and 2. Since some attention was paid in [MaT1-MaT3] to the relations of a bone $B_s$ to the two bones corresponding to periodic orbits obtained from those in $B_s$ by a period doubling bifurcation, Figure 1.4 shows how such triples of bones fit in the emanation discussion. In these figures, heavy black lines correspond to bones or to the lines a=1 and b=1, light black lines correspond to ligaments. The grey region in Figure 1.2 is the domain where the periodic orbit with maximal itinerary $(AK^{\rho(A)}BC^{-\rho(B)})^\infty$ and minimal itinerary $(BC^{-\rho(B)}AK^{\rho(A)})^\infty$ is an attractor. To simplify the annotation of the figures, we write $WQ^+$ for $WQ^{\rho(W)}$ and $WQ^-$ for $WQ^{-\rho(W)}$. In Figures 1.1, 1.2 and 1.3, if the grey lines are contracted to points (without creating further crossings on the black lines) one gets singular simplices which are embeded in the plane as would conjecturally be the corresponding sets of bones and ligaments for a two parameter family of cubic maps (see [DGMT]). In Figure 1.1, only the interior of the segment a+b=1 should be contracted to a point; in all cases, the contraction should take place without affecting the underlying bone. Figures 2.1-2.4 are counterparts respectively to Figures 1.1-1.4, obtained numerically from the cubic family given by $f_{a,b}(x)=x^3-ax+b$. Figure 0 showed for this family the region of its parameter space that corresponds most closely with **P**$^+$: in the region shown in that figure, the maps are all bimodal except at the $\alpha$-seed (a=b=0), and there is an invariant interval containing both critical points.

**Proof of Theorem 3:**

Maximal kneading sequences are monotonously increasing with a, for fixed b, and with b, for fixed a. Similarly, minimal kneading sequences are monotonously decreasing with a, for fixed b, and with b, for fixed a. Consequently, the topological entropy of $T_{a,b}^+$ increases when any of the parameters increases, the other parameter being fixed. Hence, the topological entropy on a ligament is minimal at its internal endpoint. On a bone $B_s=B_s^C \cup B_s^K$, the topological entropy of $T_{a,b}^+$ is minimal at the corners of $B_s^C$ and $B_s^K$ which are closest to the line a+b=1. Since up to a finite number of periodic orbits, the periodic orbit structure of $T_{a,b}^+$ is unchanged on the ligament part of each bone (they correspond to lines in a hyperbolic region in the polynomial case), the topological entropy is unchanged there, and can only change by increasing when following a non-ligament part parallel to either the a-axis or the b-axis. This shows that at $\alpha$, $\chi$ and $\psi$ points, the topological entropy is not greater than on the lines emanating there, which is what we had to prove.

∎

Combining Theorems 2 and 3 with the elementary properties of the bones and ligaments in $\mathbf{P}^+$ yields the following result:

**Theorem 4:** In $\mathbf{P}^+$, any two seeds can be joined by finite path made of pieces from finitely many bones and ligaments.

# Figure Captions

**Figure 0**     Loci of some finite kneading sequences for a two-parameter cubic family.

**Figure 1.0**   Graph and parameter space of the stunted sawtooth family, $T^+_{a,b}$.

**Figure 1.1**   The $\alpha$-seed in the family $T^+_{a,b}$.

**Figure 1.2**   A $\chi$-seed in the family $T^+_{a,b}$.

**Figure 1.3**   Two $\psi$-seeds in the family $T^+_{a,b}$.

**Figure 1.4**   Period-doubling in the family $T^+_{a,b}$.

**Figure 2.1**   The $\alpha$-seed in the cubic family $f_{a,b}$.

**Figure 2.2**   A $\chi$-seed in the family $f_{a,b}$.

**Figure 2.3**   Two $\psi$-seeds in the family $f_{a,b}$.

**Figure 2.4**   Period-doubling in the family $f_{a,b}$.